\documentclass[10pt,aps,twocolumn,superscriptaddress]{revtex4-2}
\makeindex
\makeatletter

\renewcommand*{\p@subsection}{}

\renewcommand*{\p@subsubsection}{}
\makeatother
\usepackage{lipsum}  
\usepackage{graphicx}
\usepackage{multirow}
\usepackage{url}
\usepackage{amsmath}
\usepackage{siunitx}
\usepackage{float}
\usepackage[none]{hyphenat}
\usepackage{lipsum}
\usepackage{color,soul}
\usepackage{xcolor}

\begin{document}

\title{Defining graphenic crystallites in disordered carbon: moving beyond the platelet model}

\author{K. J. Putman}
\affiliation{Department of Physics and Astronomy, Curtin University, Perth, Australia}
\email[Mail: ]{kate.putman@curtin.edu.au}

\author{N. A. Marks}
\affiliation{Department of Physics and Astronomy, Curtin University, Perth, Australia}

\author{M. R. Rowles}
\affiliation{John de Laeter Centre, Curtin University, Perth, Australia}

\author{C. de Tomas}
\affiliation{Department of Physics and Astronomy, Curtin University, Perth, Australia}

\author{J. W. Martin}
\affiliation{Department of Physics and Astronomy, Curtin University, Perth, Australia}

\author{I. Suarez-Martinez}
\affiliation{Department of Physics and Astronomy, Curtin University, Perth, Australia}

\date{\today}

\begin{abstract}

We develop a picture of graphenic crystallites within disordered carbons that goes beyond the traditional model of graphitic platelets at random orientation. 
Using large atomistic models containing one million atoms, we redefine the meaning of the quantity $L_{\mathrm{a}}$ extracted from X-ray diffraction (XRD) patterns.
Two complimentary approaches are used to measure the size of graphenic crystallites, which are defined as regions of regularly arranged hexagons.
Firstly, we calculate the X-ray diffraction pattern directly from the atomistic coordinates of the structure and analyse them following a typical experimental process. Second, the graphenic crystallites are identified from a direct geometrical approach.
By mapping the structure directly, we replace the idealised picture of the crystallite with a more realistic representation of the material and provide a well-defined interpretation for $L_a$ measurements of disordered carbon.
A key insight is that the size distribution is skewed heavily towards small fragments, with more than 75\% of crystallites smaller than half of $L_{\mathrm{a}}$.

\end{abstract}

\maketitle

\section{Introduction}

Disordered carbons are a versatile collection of materials with significant industrial applications such as batteries, carbon molecular sieves and super-capacitors \cite{dasgupta_2017, bianco_2020}. 
Disordered carbons span a variety of structural forms including porous carbon, glassy carbon and carbon blacks \cite{robertson_1986, harris_2005}. 
Unlike an amorphous material, disordered carbons exhibit short-range order with networks of regularly arranged hexagons; in this work we refer to these regular hexagonal regions as graphenic crystallites. 
Accurate assessment of this nanostructure is crucial for disordered carbon applications such as in batteries where charge mobility is highly dependent upon crystallite size and distribution \cite{inagaki_carbon_2010, frackowiak_2001}. 

X-ray diffraction (XRD) was the first approach to show that disordered carbons are not completely amorphous but possess short-range order \cite{berl_1932, hofmann_1931, warren_1934}. 
The disordered carbon XRD pattern contains three significant diffuse peaks: the (002) reflection from stacking, and the (10) and (11) in-plane reflections from graphenic order. 
From the (002) reflection we obtain $L_{\mathrm{c}}$, measuring the average height of regions with regular stacking. 
Both in-plane reflections can be used to extract $L_{\mathrm{a}}$, which measures the average lateral size of graphenic crystallites. 
These two size parameters are the foundation of the platelet model, first described by Riley in 1939 \cite{riley_1939} and Biscoe and Warren in 1942 \cite{biscoe_warren_1942}.
In this model, crystallites are discrete cylindrical units with height $L_{\mathrm{c}}$ and diameter $\approx$$ L_{\mathrm{a}}$, which are randomly oriented within a carbon matrix.  
This idea was further developed by Franklin \cite{franklin_1951} who proposed cross-links that connect the platelets. 

Although the platelet model was never proposed to be universally applicable, it has nevertheless dominated interpretation of disordered carbons for decades \cite{jenkins_1976, oberlin_1984, kaneko_1992}.
The strongest criticism of the platelet model has come from Marsh and Rodriguez-Reinosor \cite{marsh_2006} who argued that the fundamental nanostructure in disordered carbons is instead a curved 3D graphenic network which may contain regions of stacking. 
In their view, $L_{\mathrm{a}}$ and $L_{\mathrm{c}}$ are decoupled and describe different regions of the overall structure. 
This distinction is supported by electron micrographs showing fringes that extend past the regions of stacking \cite{jenkins_1971}. 

Computational models of the crystallite offer a pathway to quantify $L_{\mathrm{a}}$ and $L_{\mathrm{c}}$ and to interpret their meaning. 
The first of these models (circa 1940-50) from Warren \cite{warren_x-ray_1941} and Diamond \cite{diamond_x-ray_1957} used a simplified form of the Debye scattering equation \cite{debye_zerstreuung_1915} to assess the effect of sheet size and shape on platelet-like crystallites. 
More current works \cite{fujimoto_theoretical_2003, puech_new_2019}, including works from the authors  \cite{ lim_universal_2020, putman_2021}, employ the full Debye equation to simulate in-plane diffraction as well as stacking effects such as inter-layer orientation and spacing. 
Attempts to model the non-uniform structure of disordered carbon have used idealised features such as cylindrical curvature \cite{li_2007} or spherical crystallites with a sheet-size distribution \cite{dopita_2013}. 
While these works give insight into diffraction line broadening effects of disordered carbon which produce non-unique diffraction patterns \cite{jones_measurement_1938, popa_2002}, they continue to follow the platelet model with isolated cystallites dominated by edges instead of an interconnected network.

\begingroup
\begin{figure*}
  \includegraphics[width=0.9\textwidth]{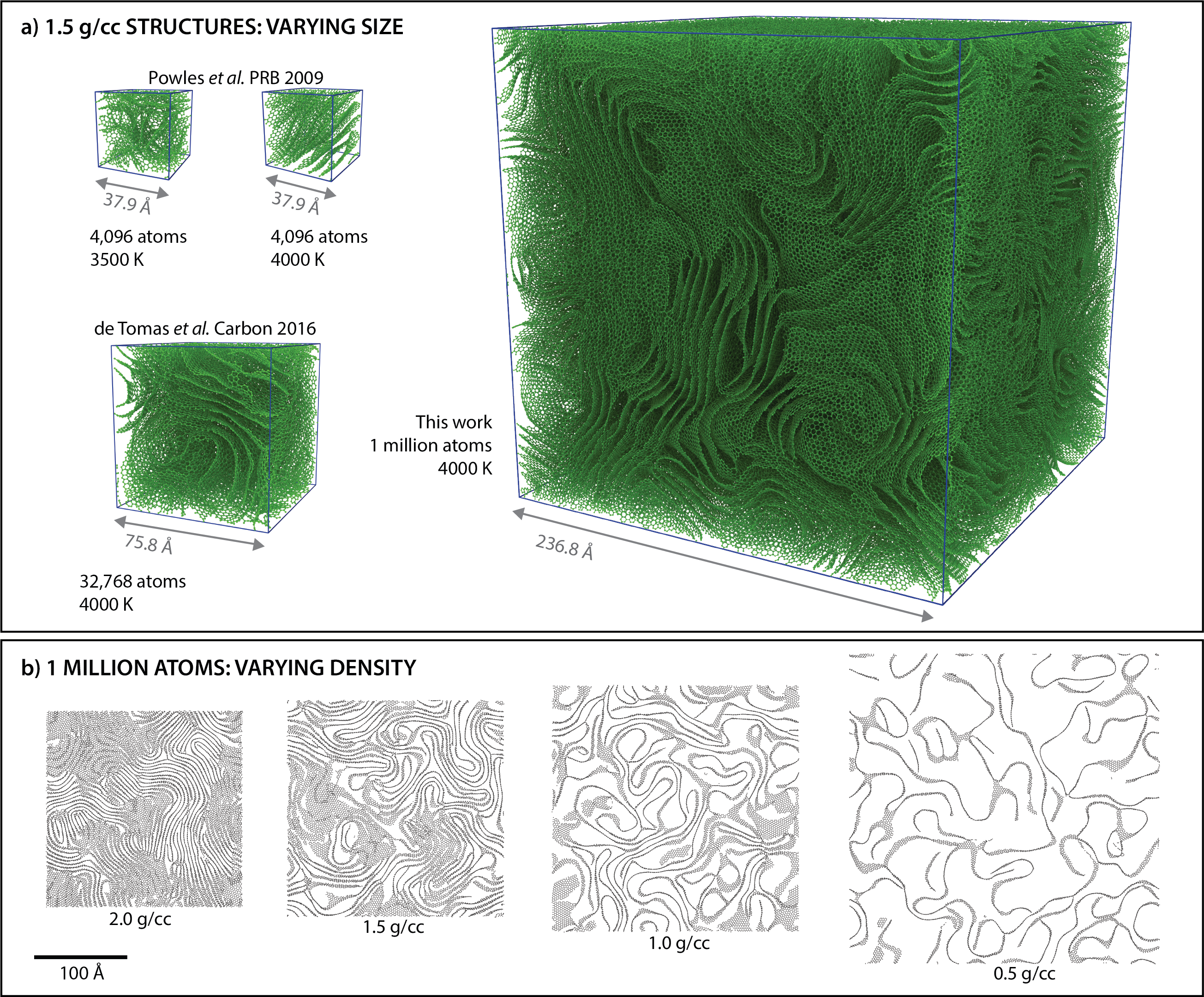}
\caption{(a) Historical series of annealed carbon structures prepared at
1.5~g/cc showing the progression of system size. Top-left shows
4,096-atom structures from Powles \emph{et al.}\cite{powles_2009} after
annealing at two different temperatures. Bottom-left shows a 32,768-atom
structure from de~Tomas \emph{et al.}\cite{detomas_2016} annealed at 4000~K.
Structure from this work (right) contains one million atoms and was annealed at
4000~K. (b) One million atom structures from this work for four densities. For clarity, cross-sections of 5~\AA\ thickness are shown.
}
\label{structures}
\end{figure*}
\endgroup

In this work we move beyond the platelet model by generating large atomistic structures of disordered carbon without imposing predefined geometries. 
The structures are generated using an annealed molecular dynamics (MD) approach \cite{deTomas_2017,deTomas_2018} which employs self-assembly to create a 3-dimensional interconnected network of curved sheets with minimal edges. 
Graphenic crystallites within the disordered carbon structures are analysed using two complimentary approaches.
First, the XRD pattern is calculated and $L_a$ is extracted from the diffraction peak profile as in an experimental measurement.
Next, the graphenic crystallites are identified by a geometric search of the structure, which reveals a predominantly circular fragment shape and a size distribution that decays exponentially. 
Comparing the geometrical size distribution with $L_a$ from the XRD pattern reveals that the largest crystallites have a size $\sim$2$L_a$, while more than 98\% of the crystallites are smaller than $L_a$. Similar behaviour is found in structures with densities spanning 0.5 to 2.0~g/cc, suggesting these trends are applicable for a wide variety of disordered carbons.

The paper is arranged as follows: Section~2 introduces the large atomistic structures, Section~3 outlines the reciprocal-space analysis of the graphenic crystallites with calculated XRD patterns, while Section~4 defines a real-space analysis where crystallites are identified geometrically. In Section~5 we discuss the implications for understanding graphenic crystallites and their intereptation in XRD.


\section{Million-atom structures}

\begin{table*}[]
\caption{Properties of the million-atom structures generated in this work.
Side-length of the supercell is listed as box size. Number of atoms in the structure is
after removal of floating clusters.  Ring statistics of the network are
computed on a per-atom basis. Coordination fractions are computed using a
cutoff of 1.85~\AA.}

\renewcommand{\arraystretch}{1.3}

\begin{tabular}{>{\centering}p{0.08\textwidth} | >{\centering}p{0.08\textwidth} | >{\centering}p{0.09\textwidth} | >{\centering}p{0.09\textwidth} >{\centering}p{0.09\textwidth} >{\centering}p{0.09\textwidth} >{\centering}p{0.09\textwidth} >{\centering}p{0.09\textwidth} | >{\centering}p{0.07\textwidth}>{\centering}p{0.07\textwidth}>{\centering\arraybackslash}p{0.07\textwidth}}
\hline 

\multicolumn{1}{c|}{Density} & 
\multicolumn{1}{c|}{Box size} &
\multicolumn{1}{c|}{Number} &
\multicolumn{5}{c|}{Rings per atom} &
\multicolumn{3}{c}{Coordination fraction (\%)}\\
(g/cc) & (\AA) & of atoms & pentagons & hexagons & heptagons & octagons & nonagons & $sp$ & $sp^2$ & $sp^3$ \\ 
\hline 
0.5 & 341.6 & 991,567  & 0.03  & 0.42  & 0.03  & 0.003  & 0.001  & 2.0  & 97.5  & 0.5  \\ 
1.0 & 271.1 & 997,561  & 0.03  & 0.43  & 0.03  & 0.003  & 0.001  & 1.7  & 97.7  & 0.6  \\  
1.5 & 236.8 & 999,535  & 0.03  & 0.42  & 0.03  & 0.003  & 0.001  & 1.6  & 97.6  & 0.8  \\ 
2.0 & 215.1 & 1,000,089 & 0.03  & 0.42  & 0.03  & 0.004  & 0.002  & 1.3  & 97.2  & 1.5 \\   
\hline \end{tabular}
\label{ring_stats}
\end{table*}

The atomistic structures were generated using a self-assembly Molecular
Dynamics (MD) approach in which we anneal the system just below the melting
point. In the self-assembly approach bonds can be made and broken, but the
structure does not diffuse like a liquid, driving the development of a high
$sp^2$ fraction and extended graphenic order. We first explored the
self-assembly approach in Ref.~\cite{powles_2009}, considering small systems
with 4096 atoms and a density of 1.5~g/cc (see Fig.~\ref{structures}a), and in
2016 expanded to 32,768 atoms at the same density \cite{detomas_2016}. 
Using these systems, we have confirmed that these
structures accurately reproduce a wide range of material properties including
microstructure, thermal conductivity \cite{suarez-martinez_2011}, 
elastic constants and pore-size distributions.
Here we extend the limits even further by increasing the system size to circa 1
million atoms. 

The MD simulations are performed in \texttt{LAMMPS} \cite{plimpton_1995}, and atomic interactions are
computed using the Environment Dependent Interaction Potential (EDIP) for
carbon \cite{marks_2000}. The initial stucture is a face-centered-cubic lattice with
63$\times$63$\times$63 unit cells (\textit{i.e.}\ 1,000,188 atoms). Using the same
approach used by ourselves to prepare carbide-derived carbons
\cite{deTomas_2017}, atoms are displaced randomly by a maximum of 0.2~\AA\
and minimized prior to annealing to remove excess potential energy.  
The temperature is maintained at 4000~K for 2~ns using the
Bussi thermostat \cite{bussi_2007} and a timestep of 0.2~fs.
At the conclusion of the annealing, the
structure is minimized to zero kelvin and any floating atoms (typically
triangles) which are not connected to the main structure are removed. A
threshold of 1.85~\AA\ is used to define a bond, and the same cutoff is
employed for computing coordination fractions and ring statistics.  The entire
procedure is repeated for densities of 0.5, 1.0, 1.5 and 2.0~g/cc, generating a
set of structures which are shown in cross-section in Fig.~\ref{structures}b.

The historical series in Fig.~\ref{structures}a demonstrates the importance of
large structures.  In the 4096-atom work \cite{powles_2009} the structure annealed at 3500~K
develops modest graphenic character with high curvature. Increasing the
temperature to 4000~K results in graphitic structure which is artificially
templated by the periodic boundary conditions. Repeating the same 4000~K
calculation with 32,768 atoms removes the templating effect, but the
crystallite size is constrained by the supercell size \cite{detomas_2016}. With the present 1~million-atom structures, all of these constraints are removed and large
crystallites can develop without any self-interaction effects.

The series in Fig.~\ref{structures}b shows a pronounced evolution in structure
with density. At the lowest density of 0.5~g/cc, large pores are present and
there is no significant stacking of graphene sheets.  As density increases the
size of pores falls and stacking begins to develop. At 1.5~g/cc, similar to the
density of glassy carbon, very few pores are present and the stacking becomes
more regular. At the highest density of 2.0~g/cc, approaching the density of
graphite, no pores are seen and large regions of extended stacking are present.
The properties of the four structures as a function of density are summarized
in Table~\ref{ring_stats}. Ring statistics are calculated using
\texttt{polypy} \cite{polypy} employing the shortest-path algorithm of
Franzblau \cite{Franzblau_1991}. Intriguingly, the ring statistics of the four structures
are basically identical, despite the striking microstructural evolution in
stacking and pore size seen in Fig.~\ref{structures}b.  The coordination
fractions are also very similar, with all four structures exhibiting a very
high $sp^2$ fraction of around 97.5\%; the only trend with density is a slight
reduction in $sp$ bonding and a concomitant increase in $sp^3$ bonding.  The
graphenic nature of the network is reflected in both the high $sp^2$ fraction
and the large number of hexagons; for comparison, graphene has 0.5
hexagons/atom with the slighly small number of $\approx$0.42 hexagons/atom 
observed here reflecting other ring types associated with line defects and 
curvature.

The properties listed in Table~\ref{ring_stats} compare favourably to
structures from similar calculations using the more sophisticated GAP
machine-learning potential \cite{wang_2022}. Although the system size and annealing temperature
are slightly smaller than our work (131,072 atoms and 3500~K, respectively) the
structural evolution is almost exactly the same. For the GAP structures the
$sp^2$ fraction is near-constant at around 97\%, the $sp$ fraction is around
1--3\% and the $sp^3$ fraction is 0.4--0.9\%. As in our work, the $sp$ fraction
drops with density, while the $sp^3$ fraction slightly increases. Similar
agreement is seen for ring statistics, with the GAP structures finding little
variation across the density range. The only difference is that the number of
hexagons with GAP is smaller at 0.30 rings/atom which is in keeping with the
lower annealing temperature of 3500~K. This general agreement between our EDIP
calculations and the more sophisticated GAP method provides reassurance that
our structures represent an accurate model of disordered carbon.


\section{Reciprocal Space analysis}

\begin{figure}
  \includegraphics[width=0.48\textwidth]{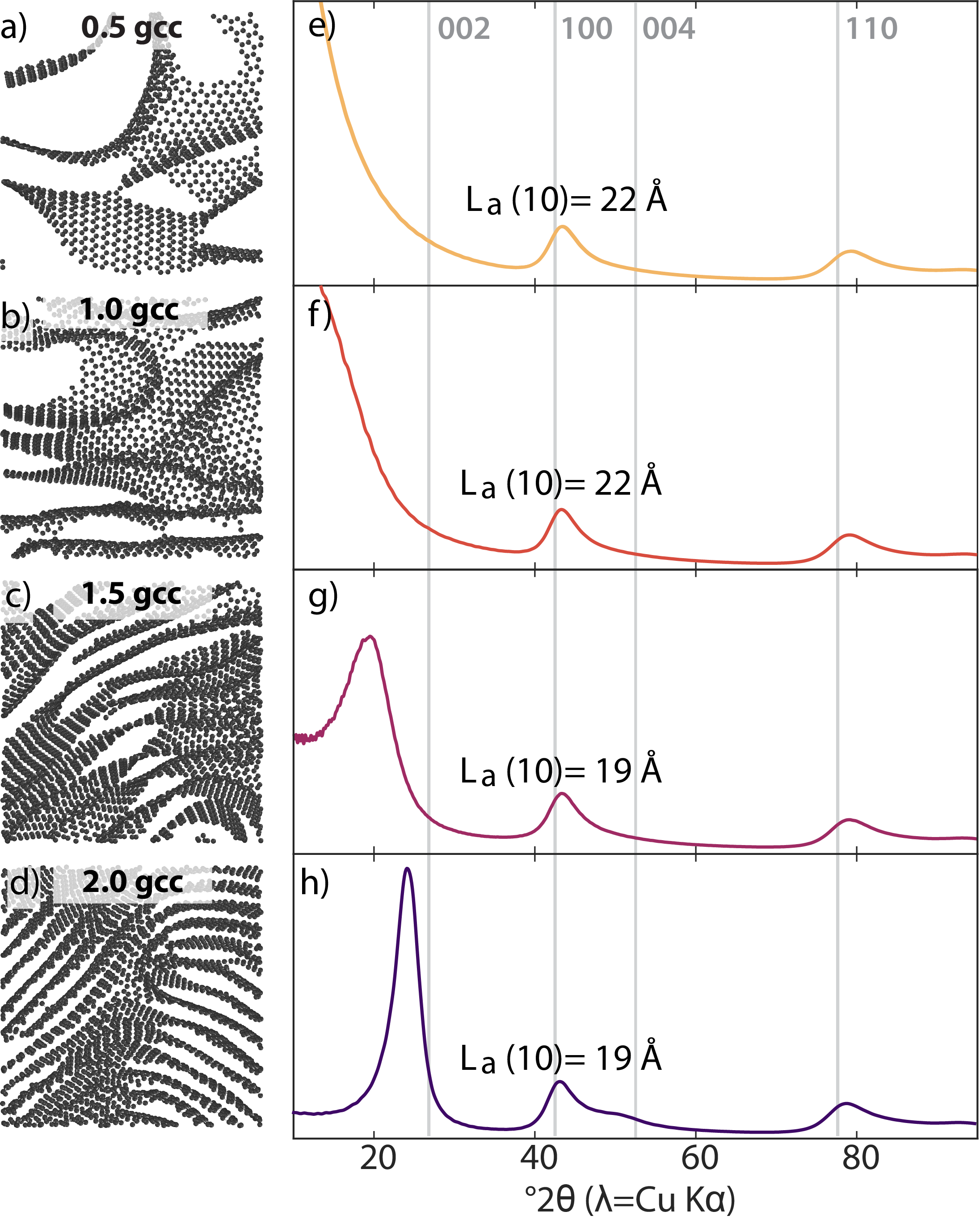}
  \caption{(a-d) Portions of the self-assembled models. Portions are 40$\times$40$\times$5~\AA$^3$. (e-f) XRD patterns of the patterns from (a-d). Vertical lines are included for reference, showing the Bragg reflections for graphite. 
Crystallite size $L_{\mathrm{a}}(10)$ is calculated from Eq.~\ref{eqn:scherrer} using the (10) reflection at  $2\theta$$\simeq$43$^{\circ}$.
  }
\label{scans}
\end{figure}

The reciprocal space diffraction pattern is calculated for each of the atomistic structures. The pattern is then analysed following a typical experimental process. 
Diffraction are computed using \texttt{debyer} \cite{debyer}, which employs a simplified monatomic form of the Debye equation \cite{debye_zerstreuung_1915},

\begingroup
\begin{equation}
\label{eqn:debye}
\begin{gathered}
I(Q) \approx \frac{1}{N} f(Q)^2 \left(\ N+2 \sum_{k=1}^{N_{\mathrm{bins}}} \frac{ \sin{Q r}}{Q r} \right)\
\end{gathered}
\end{equation}
\endgroup

with the scattering vector $Q={4 \pi \sin{\theta}}/{\lambda}$. Using bins with width, $w$=0.001~\AA, to catalogue interatomic distances, $r$, the diffraction intensity per atom, $I$, is computed as a histogram. Here, $N$ is the number of atoms in the structures and $f$ is the atomic scattering factor. The calculated diffraction patterns for all structures are analysed by Reitveld refinement \cite{rietveld_1969, loopstra_1969} in \texttt{TOPAS} \cite{TOPAS}, the same as a typical experimental analysis.
The full-width half-maximum, $B$, and peak position, $\theta$, are extracted from the (10) in-plane peak, in order to calculate the graphenic crystallite size ($L_{\mathrm{a}}$). 
Using the Scherrer equation \cite{scherrer_bestimmung_1918}, $L_a$ is calculated as,

\begin{equation}
\label{eqn:scherrer}
L_{\mathrm{a}}=K \frac{\lambda}{B \cos{\theta}}
\end{equation}

where $\lambda$ is the wavelength of the X-rays, and $K$ is the shape factor. 
In our previous work \cite{lim_universal_2020, putman_2021}, $K$ is derived empirically using a linear fit of $L_a$ from simulated crystallites of known size against the peak parameters $\lambda/(B \cos\theta)$. 
From this empirical fit, $K$ is equal to 0.9 for ordered stacks of three or more layers \textit{i.e.}\ 3-dimensional crystallites with symmetric diffraction peaks. 
Conversely for a single layer or disordered stacking, where the layers diffracts as a collection of 2-dimensional crystallites, $K$ is equal to 2 and Eq.~\ref{eqn:scherrer} is modified with a non zero intercept of -11~\AA. 
However, this affine form of the Scherrer equation breaks down when $L_{\mathrm{a}}$$<$20~nm. 
As the crystallites in this study are well below this size we follow the convention of Iwashita \textit{et al.} \cite{iwashita_specification_2004} where the shape factor is set to unity when it is not well defined, \textit{i.e.}\ $K$=1. 
The average height of regions with parallel stacked layers, $L_c$, is similarly obtained using Eq.~\ref{eqn:scherrer}, where $B$ and $\theta$ come from the (002) peak and $K$=0.9. 
The distance between these layers, also known as the $d$-spacing ($d_{002}$), is calculated using the Bragg equation and the average number of layers estimated to be $N=L_{\mathrm{c}} / d_{002}$ \cite{putman_2021}.

\begin{figure}[b]
  \includegraphics[width=0.48\textwidth]{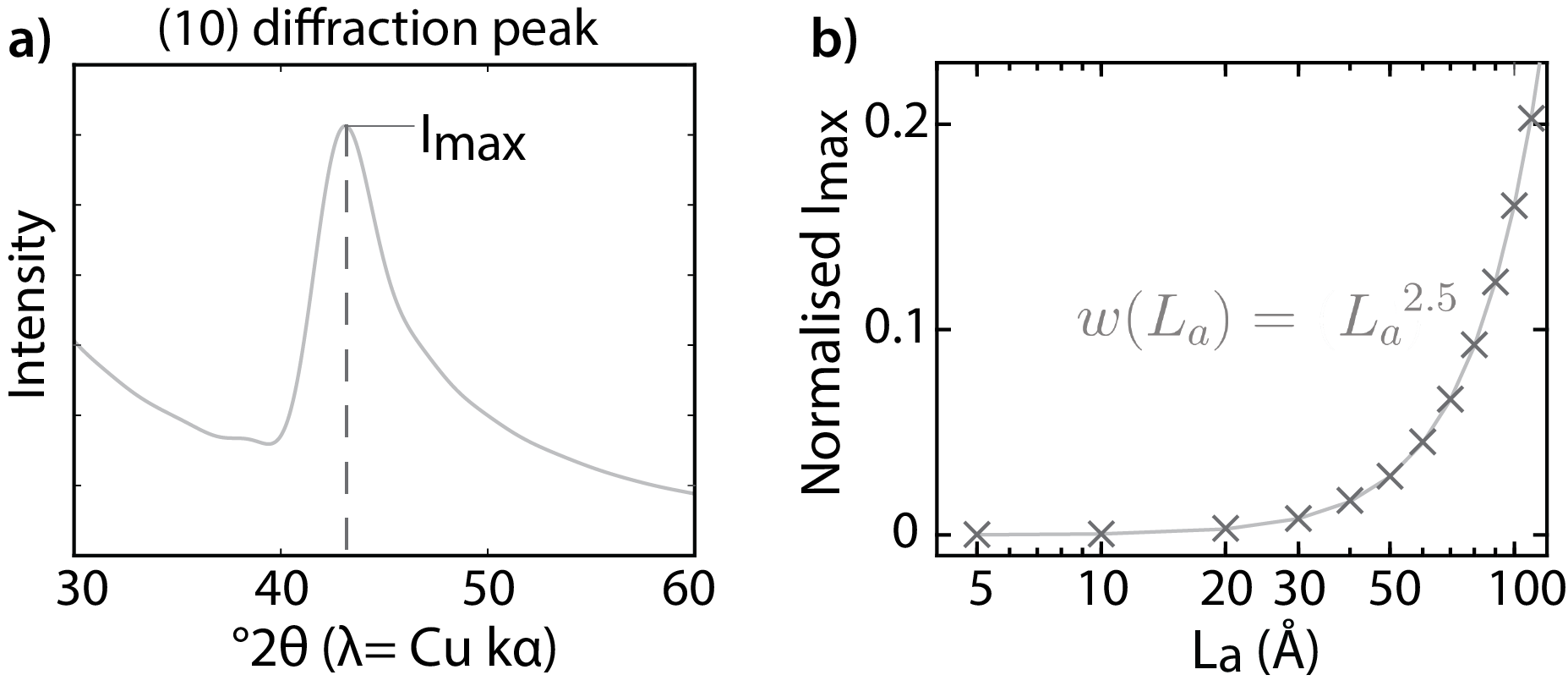}
  \caption{a) The (10) diffraction peak shown for a circular fragment for $L_{\mathrm{a}}$=50~nm, where the vertical line indicated the position and height taken as the maximum intensity, $I_{\mathrm{max}}$. b) $I_{\mathrm{max}}$ is plotted against the crystallite size, $L_{\mathrm{a}}$. This relationship is fit to the function $w(L_{\mathrm{a}})$=${L_{\mathrm{a}}}^{2.5}$, where $w$ is taken as the size dependant intensity weighting. }
\label{weighting}
\end{figure}

\begin{figure*}
  \includegraphics[width=0.99\textwidth]{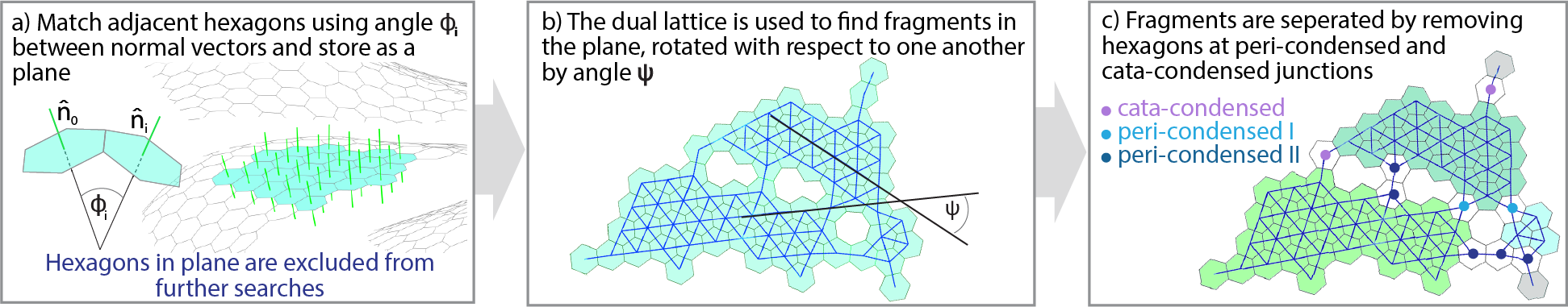}
  \caption{Workflow diagram outlining the real space geometrical approach used to extract the graphenic crystallites from the self-assembled models.}
\label{algorithm}
\end{figure*}

Figure~\ref{scans} shows cross-sections of the atomistic structures for the densities 0.5, 1.0, 1.5 and 2.0~g/cc (Fig.~\ref{scans}a-d) next to their computed diffraction patterns (Fig.~\ref{scans}e-f). 
For reference, vertical lines overlay the patterns showing the primary Bragg reflections in graphite; these are the (002) and (004) from layer stacking, and in-plane (100) and (110). 
The diffraction patterns for all structures contain two small and broad peaks which are the in-plane (10) and (11) reflections arising from graphenic structure. 
These in-plane peaks have almost identical profiles for all four densities suggesting there is little change in the graphenic crystallite size across the structures.
This is further confirmed by similar measurements of crystallite size taken from the diffraction peak, where $L_a (10)$ is between 19 and 22~\AA.

The  (002) stacking reflection is present for the two higher density models (Fig.~\ref{scans}g and h), 1.5 and 2.0~g/cc, coinciding with the appearance of significant stacking visibly evident in the atomistic structures (Fig.~\ref{scans}c and d).  
For the 1.5~g/cc model, the (002) peak sits at $2\theta \approx$19$^{\circ}$ with $L_{\mathrm{c}}$=8.3~\AA\ giving an interlayer distance of $d_{002}$=4~\AA, indicating an average of only two layers.  
While stacks with more than two layers can be seen in the atomistic structure, few of these layers are in fact parallel over a large enough region to produce coherent reflections.
In the 2.0~g/cc model, the (002) peak shifts to a higher angle while the peak profile narrows and increases in intensity so that the second order (004) reflection is also resolved at $\sim$50$^{\circ}$. 
The shift of the (002) peak between 1.5 and 2.0~g/cc models is due to a decrease of the $d_{002}$-spacing from 4 to 3.7~\AA, while the narrowing of the peak indicates a greater number of regularly stacked layers ($\sim$8). 

Additional diffraction patterns are calculated (using Eq.~\ref{eqn:debye}) in order to obtain a weighting function for the relationship between crystallite size and diffraction intensity (Fig.~\ref{weighting}). 
Diffraction patterns are calculated for a series of atomistic circular fragments of known size cut from a perfect graphene sheet (see Ref~\cite{lim_universal_2020} for further details).
Maximum intensity, $I_{\mathrm{max}}$, of the (10) peak is obtained from the diffraction pattern of each crystallite as shown in Fig.~\ref{weighting}a.
An empirical fit the normalised $I_{\mathrm{max}}$ against $L_{\mathrm{a}}$ (Fig.~\ref{weighting}b)  the gives the weighting function, $w(L_{\mathrm{a}})$.  
The exponential form of $w(L_{\mathrm{a}})$ appears to arise from two factors. 
First the numbers of atoms increases with area as ${L_{\mathrm{a}}}^2$.
Second, the scattering per atom increase with area, and following the convention of Diamond \cite{diamond_x-ray_1957}, the graphenic crystallites have size $L_{\mathrm{a}}$=$\sqrt{\mathrm{area}}$.
The combination of these two factors gives the same relationship obtained by the empirical fit, where $w(L_{\mathrm{a}})$=${L_{\mathrm{a}}}^{2.5}$.


\section{Real space analysis}

\begin{figure}[b]
\includegraphics[width=0.48\textwidth]{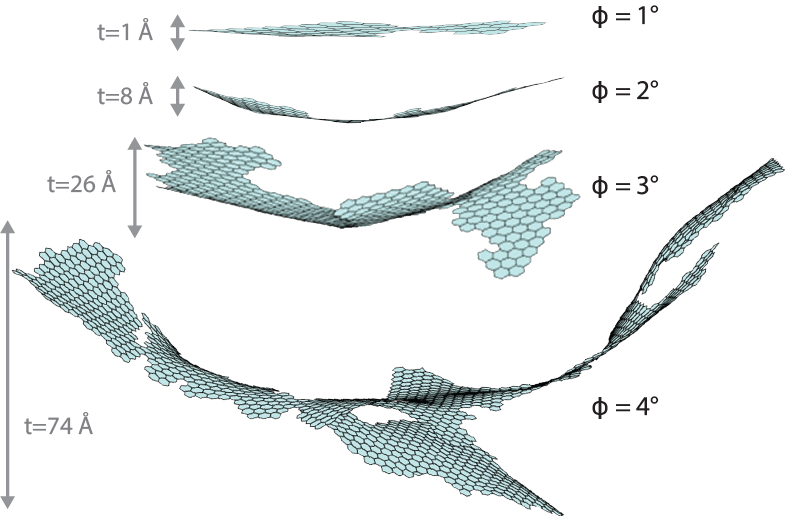}
 \caption{Crystallites identified for threshold angles $\phi$= 1 to 4$^{\circ}$. The crystallites shown have the greatest apparent thickness, $t$, as a result of curvature.}
\label{depth}
\end{figure}  

Complementary to the reciprocal space analysis, the structures are analysed in real-space by a computational process of geometrically identifiying the graphenic crystallites as outlined in Fig.~\ref{algorithm}. 
We begin with a network of hexagons obtained from the ring analysis; all non-hexagonal rings are discarded as they do not contribute to graphenic fragments.
Planes are identified in the hexagonal ring network (Fig.~\ref{algorithm}a) via a threshold angle, $\phi$.
A plane is first defined by an arbitrary hexagon, $h_0$, with normal vector, $\hat n_0$. This vector is compared to the normal vector for surrounding hexagons, $\hat n_i$ using the angle between the two, $\phi_i$; all hexagons with $\phi_i$$<$$\phi$ are added to the plane. 
This process is repeated for hexagons in the plane, with each hexagon re-defined as $h_0$, until no further matches are found. 
Hexagons added to a plane are excluded from the search and the process is repeated until all hexagons have been analysed. 
No assumptions are made \textit{a priori} about the correct value for $\phi$, so this process is replicated in integer steps of 1$^{\circ}$ from $\phi$=1$^{\circ}$ to 10$^\circ$. 

Once the planes are identified, they are subsequently separated into individual fragments via a process illustrated in Fig.~\ref{algorithm}b,c.
The dual lattice is used to find in-plane rotations ($\psi$) between adjacent clusters of hexagons and identify junctions where the plane can separated into individual crystallites.
We borrow the nomenclature cata-condensed and peri-condensed from organic chemistry to identify the junctions between individual crystallites. 
Figure~\ref{algorithm}c shows three types of junctions with this classification. 
Cata-condensed fragments are separated by removing the hexagons with two non-adjacent dual edges (Fig.~\ref{algorithm}). 
This has the added benefit of removing fragments composed of a single line of hexagons. Fragments connected by one (I) or two (II) peri-condensed hexagons are separated by identifying and removing dual vertices with two sets of non-adjacent edges. 
Finally, we remove fragments with only one or two hexagons as they are too small to be considered as crystallites. 

\begingroup
\begin{figure}
  \includegraphics[width=0.47\textwidth]{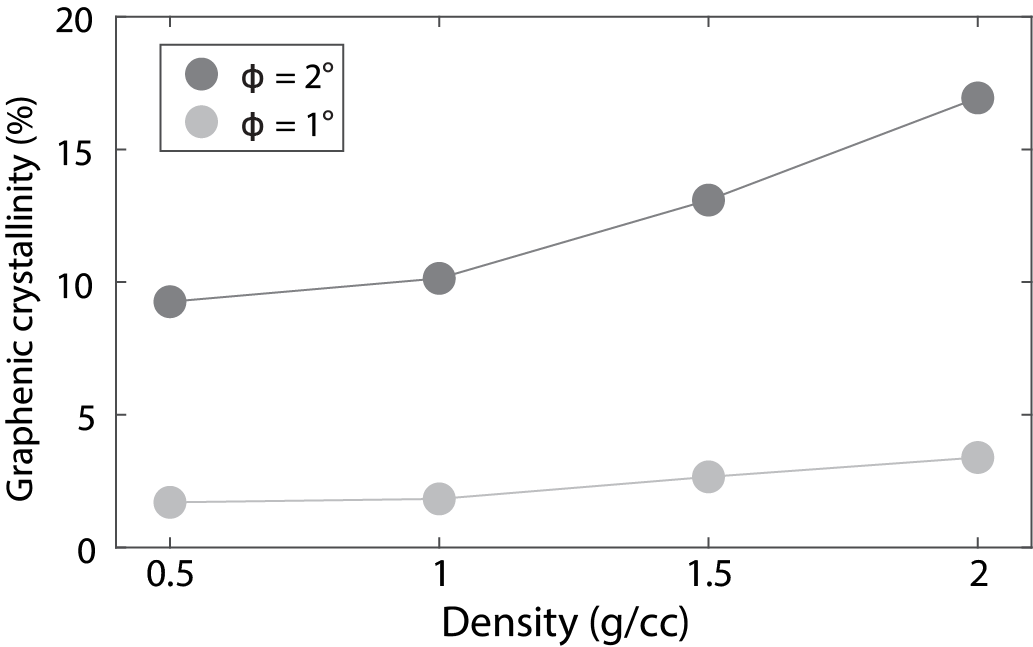}
  \caption{The graphenic crystallinity for all densities for thresholds of $\phi$=1$^{\circ}$ and 2$^{\circ}$. Graphenic crystallinity is defined as the percentage of rings contained in the crystallites with respect to the total number of rings in the structure (both hexagonal and non-hexagonal). }
\label{crystallinity}
\end{figure}
\endgroup

\begingroup
\begin{figure}
  \includegraphics[width=0.52\textwidth]{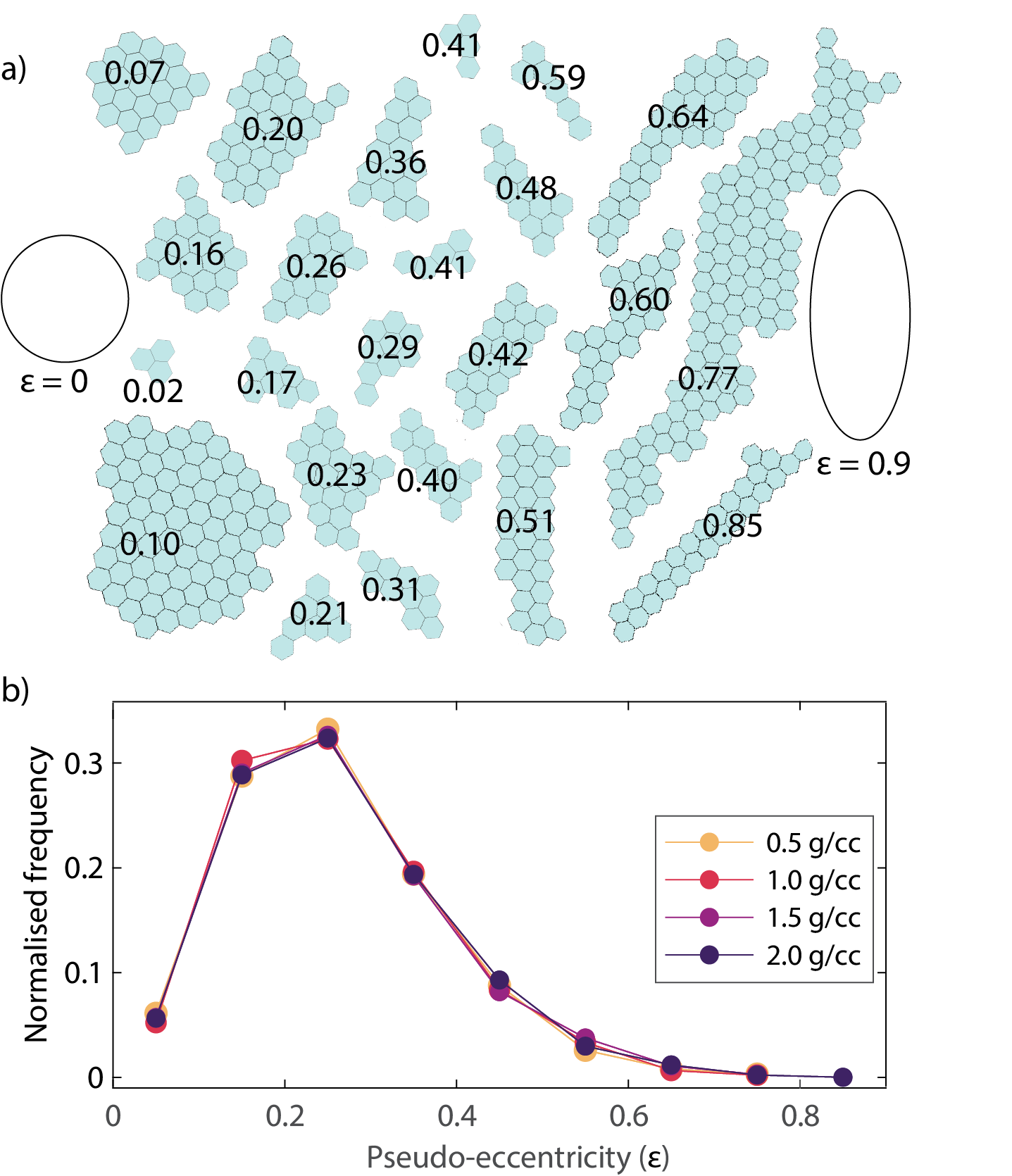}
  \caption{Pseudo-eccentricity, $\varepsilon$, is calculated as the standard deviation of diameters measured at intervals around the circumference of the crystallite and normalised by $L_{\mathrm{a}}$. a) Examples of crystallites labelled with their corresponding $\varepsilon$ and arranged from the circular, \textit{i.e.}\  $\varepsilon$=0, to the most elliptical crystallites with a maximum of $\varepsilon$=0.85. b) Shape distribution of $\varepsilon$ for crystallites extracted using a threshold angle of $\phi$=2$^{\circ}$.
}
\label{shape}
\end{figure}
\endgroup

Threshold angles larger than $\phi$=2$^{\circ}$ produce fragments with large curvature as seen in Fig.~\ref{depth}.
The curvature is quantified by rotating a crystallite to align the average normal vector with the $z$-axis. 
This allows the apparent thickness, $t$, of a crystallite to be measured as seen in Fig.~\ref{depth}. 
For $\phi$=4$^{\circ}$ and 3$^{\circ}$ the algorithm identifies large fragments with significant curvature and overlapping segments. 
For $\phi$=2$^{\circ}$, some curvature is still present though the resulting fragments have $t$ less than 10\% of the particle diameter. 
Conversely threshold angles of $\phi$$<$1$^{\circ}$ can be disregarded by comparing the crystallites obtained at 1$^{\circ}$ and 2$^{\circ}$. 
We identify specific crystallites for $\phi$=2$^{\circ}$ which are large and flat ($t$$<$1.5~\AA) but are separated into smaller fragments for $\phi$=1$^{\circ}$. 
Thus, the ideal threshold angle is between 1 and 2$^{\circ}$.

\begingroup
\begin{figure*}
  \includegraphics[width=0.95\textwidth]{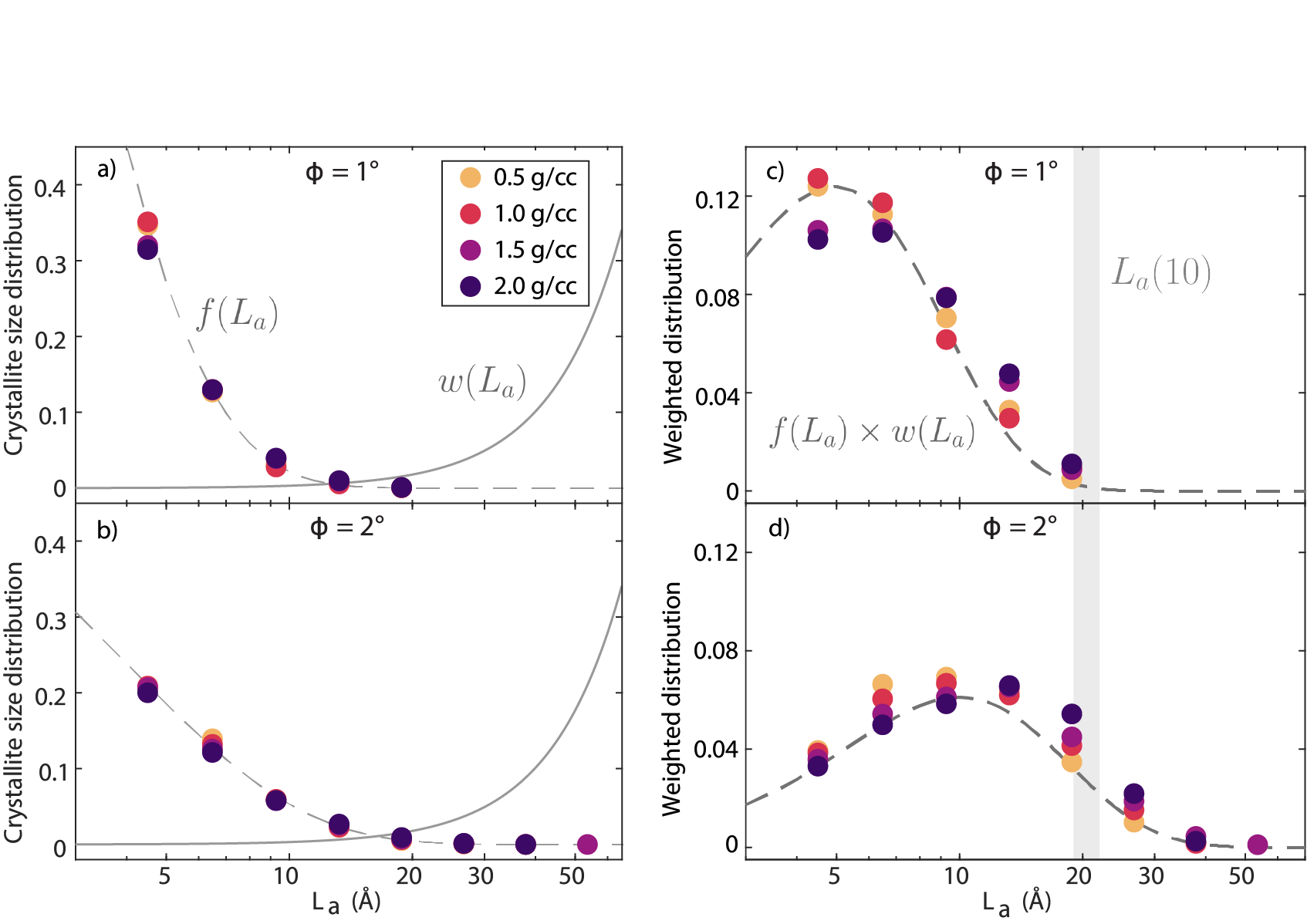}
  \caption{a-b) Crystallite size ($L_{\mathrm{a}}$) distributions for $\phi$=1$^{\circ}$ and 2$^{\circ}$. The dotted lines shows an exponential function, $f(L_a)$, fit to the data for a given angle, $\phi$. Solid grey lines show the weighting function, $w(L_a)$, for reference. c-d) The size distributions weighted by the relative intensity using $w(L_a)$. The dotted line shows the result of multiplying the weighting function and distribution fit function. Data points show the result of directly multiplying each point from the distributions by their relative weighting. The vertical band from $19-22$~\AA\ shows $L_{\mathrm{a}}(10)$ from the reciprocal space analysis.
}
  \label{dist}
\end{figure*}
\endgroup

The fraction of graphenic crystalline material in the structures is calculated as the total number of rings within the crystallites divided by the total number of rings in the structure, both hexagonal and non-hexagonal. Figure~\ref{crystallinity} shows the crystallinity for threshold angles of $\phi$=1 and 2$^{\circ}$. 
The clearest effect of the choice of $\phi$ is that the graphenic crystallinity increases by a factor of around 2. 
For either threshold angle, the graphenic crystallinity shows a slight increase with density.
The increasing graphenic crystallinity is unexpected given the calculated diffraction patterns in Figure~\ref{scans}.  
The maximum peak intensity is expected to increase with the percentage of crystallites, while the (10) diffraction peaks in Figure~\ref{scans} have essentially the same maximum intensities across the four densities.
This apparent inconsistency may be due to the small percentage of graphenic crystallites within the structures; while the relative proportion of graphenic crystallites increases with density, the absolute percentage of crystalline material remains small, representing a minority phase. 

An analysis of the crystallite shape is shown in Fig.~\ref{shape} using the pseudo-eccentricity, $\varepsilon$. Pseudo-eccentricity is computed by taking the standard deviation of a crystallite diameter, $d$, at a number of intervals around its circumference and normalising by crystallite size, \textit{i.e.}\ $\varepsilon$=$\sigma_{\mathrm{mean}}/L_{\mathrm{a}}$.  
Using this definition, a pseudo-eccentricity of $\varepsilon$=0 corresponds to a circle and $\varepsilon=0.9$ corresponds to an ellipse as shown in Fig.~\ref{shape}a. 
The distribution of $\varepsilon$ is shown in Fig.~\ref{shape}b, for $\phi$=2$^{\circ}$, though it should be noted that the distribution does not vary significantly with $\phi$. 
The majority of crystallites have a small pseudo-eccentricity,  $0.1<\varepsilon<0.3$, and are essentially circular fragments. 

Finally, the size distribution for crystallites identified by the algorithm is computed for threshold angles of $\phi$=1$^{\circ}$ and 2$^{\circ}$  (Fig.~\ref{dist}a and b respectively). 
These distributions follow an exponential decay and fit to a function of the form $f(L_{\mathrm{a}})$=$A_1 \exp{(-L_a/A_2)}$.
For $\phi$=1$^{\circ}$, the proportion of crystallites comprised of three hexagons ($L_{\mathrm{a}}$$\sim$3.5~\AA) is large. 
This is not surprising given that a smaller $\phi$ separates fragments which will otherwise be identified as one larger crystallite. 
The maximum crystallite size for these distributions at this threshold are only  $\sim$17~\AA\ which is smaller than the average crystallite size $L_{\mathrm{a}}(10)$ obtained from the reciprocal space analysis (shown with a vertical bar Fig.~\ref{dist} c and d).
For $\phi$=2$^{\circ}$, the distribution shifts so that the maximum crystallite size extends to $\sim$53~\AA\ and the proportion of small crystallites decreases.
Figure~\ref{dist}c and d show the size distributions weighted by $w(L_a)$, adjusting for the diffraction intensity as a function of crystallite size (Fig.~\ref{weighting}).
The dotted line in Fig.~\ref{dist}c and d represents the of product the distribution and weighting functions. 
Applying $w(L_a)$ allows for a more intuitive way to compare the real space distribution to $L_{\mathrm{a}}(10)$.
For $\phi$=1$^{\circ}$ (Fig.~\ref{dist}c), the maximum of the weighted distribution is at $L_{\mathrm{a}}$$\approx$5~\AA\ and sits lower than the arithmetic mean of $\sim$5.7~\AA. 
For $\phi$=2$^{\circ}$ (Fig.~\ref{dist}d), the peak of the weighted distribution shifts to a larger crystallite size of $10-15$~\AA\ which is higher than the arithmetic mean of $\sim$7~\AA. For simplicity, further discussion will focus on the result obtained from $\phi$=2$^{\circ}$.

Measuring $L_{\mathrm{a}}$ experimentally is a simple process but it is unclear how the numerical value correlates to the size distribution of actual crystallites within the material.
We map $L_{\mathrm{a}}(10)$ to the actual crystallites by computing the size distribution of the graphenic crystallites.

The distribution is exponential and extends from the lower limit of 3.5~\AA\ up to around 40~\AA. For comparison, the reciprocal space analysis yields a value centred on a narrow range with $L_{\mathrm{a}}(10)$=$19-22$~\AA.
By definition $L_{\mathrm{a}}(10)$ is a measure of the average crystallite size so it is surprising that it is more representative of the small portion of larger fragments.
In fact, less than 2\% of crystallites have $L_{\mathrm{a}}$$>$19~\AA.
Even when the distribution is weighted to account for the great diffraction intensity of larger fragments (Fig.~\ref{dist}c and d) we find that $L_{\mathrm{a}}(10)$ is larger than the distribution maxima at $\sim$10~\AA. 

\begingroup
\begin{figure*}
  \includegraphics[width=0.98\textwidth]{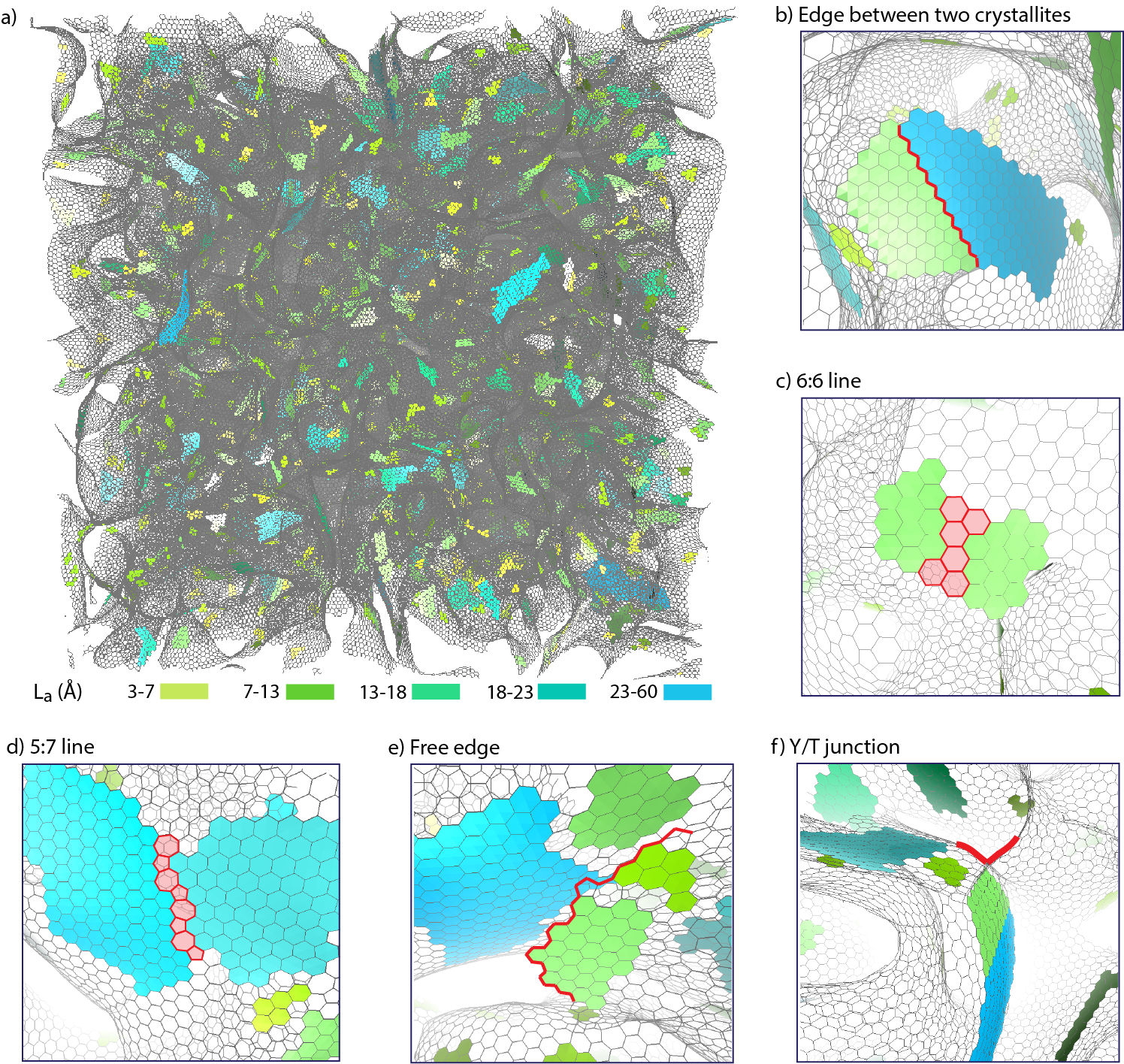}
  \caption{a) One million atom structure of density 0.5~g/cc with crystallites coloured for size. Panels b-f) show crystallites with surrounding structural features.}
\label{fragments}
\end{figure*}
\endgroup


\section{Discussion}

Our analysis allows us to develop a picture of the graphenic crystallites which differs from the platelet model and includes the local structure. 
The large structures of 1~million atoms allow features to develop which are unable to form in smaller simulation cells.
Figure~\ref{fragments}a shows the 0.5~g/cc structure with the graphenic crystallites coloured by size. 
In contrast to the traditional picture of isolated crystallites in an undefined matrix, we instead find that graphenic crystallites are incorporated into continuous sheets of hexagons with curvature as the primary grain boundary. 
Other features separating the graphenic crystallites are shown in Fig.~\ref{fragments}e-f which highlight the local structural motifs.
Two crystallites with a shared edge can be identified on sheets with cylindrical curvature, and an example with a zigzag edge is shown in Fig.~\ref{fragments}b.
In some cases the local curvature is sufficiently great so that the crystallites are separated by line of hexagons, labelled as a 6:6 line, that do not belong to either fragment (Fig.~\ref{fragments}c). 
Conversely, there are 5:7 lines of alternating pentagons and heptagons (Fig.~\ref{fragments}d) which separate crystallites with in-plane misalignment and these essentially denote the grain boundaries in the sheets.
Figure~\ref{fragments}e and f show the dominant motifs containing $sp$ and $sp^3$ bonded atoms. An armchair edge is seen in Fig.~\ref{fragments}e with alternating $sp$ and $sp^2$ bonds.
The $sp^3$ bonds primarily form Y/T junctions as seen in Fig.~\ref{fragments}f with one sheet terminating against another perpendicular sheet. 
The reduction of the $sp$ fraction in higher density structures (Table~\ref{ring_stats}) indicates that these have fewer free edges, some of which have converted into Y/T junctions.

For the 0.5~g/cc structure shown in Fig.~\ref{fragments}a, graphenic crystallites are randomly distributed throughout the structure with no preferred orientation. 
The 1.5 and 2.0~g/cc structures contain regions of stacking and one might expect the graphenic crystallites to be bounded within these regions.
Instead the crystallites are randomly distributed regardless of stacking and graphenic crystallites are present outside of the stacking regions.
Similarly the stacked regions do not need to contain only the graphenic crystallites.
The non-hexagonal rings which are not part of the graphenic crystallites can be incorporated into layers which produce stacking reflections.
Thus we observe that the in-plane ordering is not bounded by the stacking as the platelet model suggests.

The crystalline structure we define here is an important part of how we understand the overall structure of disordered carbons. 
The platelet model is derived from an understanding of how graphite is structured, as graphenic sheets stacked upon one another. 
Modelling disordered carbon from the perspective of graphite, using platelet-like crystallites, infers an overall structure of small graphite-like units.
Here, we go beyond the platelet model showing a structure of large continuous sheets with regions of high in-plane order \textit{i.e.}\ the graphenic crystallites.
With this comes a new mental picture of the crystalline order.
The graphenic crystallites exist separately to the stacking and similarly the stacking order can contain layers which are not graphenic crystallites. 
We have two distinct types of crystalline order; the in-plane ordering of graphenic crystallites and the out-of-plane ordering of layer stacking.

The size distributions in Fig.~\ref{dist} are a useful tool for interpreting an experimentally obtained measurement of $L_{\mathrm{a}}(10)$. 
By definition, $L_{\mathrm{a}}(10)$ is an average size across the crystallites, but the actual size distribution is unknown or assumed to be a log-normal or gamma distribution \cite{langford_2000}. 
From the reciprocal space analysis we find $L_{\mathrm{a}}(10)$ is between 19 and 22~\AA\ and this is at least double the arithmetic mean of the crystallite size ($\sim$7~\AA). 
While it is surprising that $L_{\mathrm{a}}(10)$ appears to misrepresent the actual crystallites, the intensity weighting applied in Fig.~\ref{dist}d offers some explanation for the difference. 
The weighting $w(L_{\mathrm{a}})$ shows that larger crystallites dominate the diffraction peak profile.
It is useful then to clarify what an experimental measurement of $L_{\mathrm{a}}$ tells us about the graphenic crystallites in a disordered carbon.
With $L_{\mathrm{a}}(10)$ situated at the right tail of the distribution it is larger than over 98\% of the actual crystallites identified: in fact, more than 75\% of crystallites are much smaller with $L_{\mathrm{a}}$$<$$\frac{1}{2}$ $L_{\mathrm{a}}(10)$.
The computed size distributions provide a unique solution to the in-plane diffraction peak of disordered carbon and the measurement of the graphenic crystallites.
It is important to note that our graphenic crystallite model is derived from structures which contain significant curvature and this is key to the shape of crystallite size distribution which extends from the smallest size of crystallite.
Experimentally, this interpretation of a disordered carbon XRD pattern should be coupled with a technique such as transmission electron microscopy (TEM) or absorption techniques to confirm the presence of pores and the associated curvature in the structure.


\section{Conclusion}

From our analysis of large atomistic structures we define the graphenic crystallites in disordered carbon.
These crystallites are not confined to isolated regions of stacking but are instead incorporated into continuous sheets throughout the material with no preferred orientation. Our key results can be summarised as follows:
\begin{itemize}
	\item The crystallite size obtained from XRD, $L_{\mathrm{a}}(10)$, marks a point in a size distribution which is roughly half the $L_{\mathrm{a}}$ largest fragment.
	\item $L_{\mathrm{a}}(10)$ is larger than that of over 98\% of the crystallites and more than 75\% of crystallites are smaller than $\frac{1}{2}$ $L_{\mathrm{a}}(10)$. 
	\item The crystallites are predominantly circular with $\sim73$\% having a pseudo-eccentricity of $\varepsilon<0.3$.
\end{itemize}

Our model of graphenic crystallite and disordered carbon structure has two primary implications for how the structure of disordered carbon is investigated experimentally. First, we provide a unique solution to the in-plane X-ray diffraction peak and a clear interpretation of $L_{\mathrm{a}}(10)$. Second, our model describes a material with significant curvature thus the XRD measurements must complemented by additional experiemental techniques to confirm the presense of pores.

\section{Acknowledgement}
ISM acknowledges fellowship FT140100191 from the Australian Research Council.
Computational resources are provided by the Pawsey Supercomputing Centre with funding from the Australian Government
and the Government of Western Australia.
JWM acknowledges funding from the Forrest Research Foundation Fellowship

\bibliographystyle{unsrt}
\bibliography{references}

\end{document}